\documentclass[11pt]{article}
\usepackage{amsfonts}
\usepackage{latexsym,amsmath}
\usepackage{amssymb,array}
\makeatletter \oddsidemargin 0in \evensidemargin 0in \textwidth
16cm \RequirePackage[dvips]{graphicx} \textheight 20cm
\newtheorem{t1}{Theorem}[section]

\begin{document}
\title{Trends and Practices in Process Capability Studies}
\author{Mahendra Saha
\footnote{Corresponding author. e-mail:
mahendrasaha@curaj.ac.in}\\Department of Statistics\\Central University of Rajasthan \\Bandarsindri, 305 817, India \and Sudhansu S.
Maiti\\Department of Statistics\\Visva Bharati
University\\Santiniketan, 731 235, India}
\date{}
\maketitle
\begin{abstract}
Quantifying the ``capability'' of a manufacturing process is an important initial step in any quality improvement program. Capability is usually defined in dictionaries as ``the ability to carry out a task, to achieve an objective''. Process capability indices(PCIs) is defined as a combination of materials, methods, equipments and people engaged in producing a measurable output. PCIs which establish the relationships between the actual process performance and the manufacturing specifications, have been a focus of research in quality assurance and process capability analysis. Capability indices that qualify process potential and process performance are practical tools for successful quality improvement activities and quality program implementation. As a matter of fact, all processes have inherent statistical variability, which can be identified, evaluated and reduced by statistical methods. Generalized Process Capability Index, defined as the ratio of proportion of specification conformance (or, process yield) to proportion of desired (or, natural) conformance. We review the process capability indices in case of normal, non-normal, discrete and multivariate process distributions and discuss the inferential aspects of some of these process capability indices. Relations among the process capability indices have also been illustrated with examples. Finally we also consider the process capability indices using conditional ordering and transforming multivariate data to univariate one using the concept of structural function.
\end{abstract}
\section{Introduction}
Process capability is an important concept for industrial managers to understand. The challenge in today's competitive markets is to be on the leading edge of producing high quality products at minimum costs. This cannot be done without a systematic approach and this approach is contained within what has been called ``Statistical quality control''. The segment of statistical quality control(SQC) discussed here is the process capability study. Process capability indices(PCIs) aim to quantify the capability of a process of quality characteristic to meet some specifications that are related to a measurable characteristic of its produced items. These specifications are determined through the lower specification limit ($L$), the upper specification limit ($U$) and the target value (T). A variety of such indices have been developed in last two decades. Following the work of Kane (1986), a number of articles have appeared introducing new indices and /or studying the properties of existing ones. Kaminsky et al. (1998) have given their critical comments on uses of these indices, and suggested a future measurement. Uses of process capability indices in administration and supplier certification have been discussed in Latzko (1985) and Schneider et al. (1995).  Excellent reviews on them are given by Rodriguez (1992), Kotz and Johnson (1993, 2002), and Kotz and Lovelace (1998). In addition, Spiring et al. (2003) provide an extensive bibliography on process capability indices. It appears to be a general acceptance of the idea that PCIs can be used only after it has been established that a process is in ``statistical control" (for example, by the use of control charts). This is reasonable if it simply requires that there be no irregular changes in quality level. However, there seems to be, in some quarters, an assumption that the measured characteristic should have a normal distribution (at least approximately), although it is difficult to see why a good industrial process must result in a normal distribution for every measured characteristic. The majority of the process capability indices discussed in the literature are associated only with processes that can be described through some continuous distributions of the characteristics and, in particular, normally distributed characteristics. The most widely used such indices are $C_{p}$  [by Juran (1974)], $C_{pk}$  [by Kane (1986)], $C_{pm}$ [by Hsiang  and Taguchi (1985)] and $C_{pmk}$ [by Choi and Owen (1990), Pearn et al. (1992)] and their generalizations for non-normal processes suggested by Clements (1989), Constable and Hobbs (1992), Pearn and Kotz (1994-1995),  Pearn and Chen (1995) and Mukherjee (1995). A number of new approaches to process capability analysis have been attempted by Carr (1991) and Flaig (1996). Another index that is given by Boyles (1994) is also worth mentioning. Often, however, one is faced with processes described by a characteristic whose values are discrete. Therefore, in such cases none of these indices can be used. The indices suggested so far whose assessment is meaningful regardless of whether the studied process is  discrete or continuous are those suggested by Yeh and Bhattacharya (1998), Borges and Ho (2001), Perakis and Xekalaki (2002, 2005) and Maiti et al. (2010).\\
In this article, our objective is to look into the overview of successive development of the process capability indices from the very beginning and also the inferential aspects of these capabilities, not only in case of normal process distribution, but also for non-normal and in some cases for discrete process distributions. The article has been organized as follows. Section $2$ deals with some notations, where as Section $3$ consider the background of process capability indices which have been developed in the earlier stage. Section $4$, $5$ and $6$ consider the same in case of non-normal process distribution, any general process distribution, whether the process is normal, non-normal or discrete process distribution and multivariate process distributions respectively. Section $6$ is devoted to gives concluding remarks.
\section{ Notations}
\begin{tabular}{lcl}
$L$ &:& Lower specification limit.\\
$U$ &:& Upper specification limit.\\
$f(x)$&:&  Probability density (or mass) function of quality characteristic X.\\
$F(x)$ &:&  Cumulative distribution function of quality characteristic X.\\
$\bar F(x)$ &:&  $1-F(x).$\\
$UDL$ &:&  Upper desired limit.\\
$LDL$ &:&  Lower desired limit.\\
$\alpha_1$&:&  $P(X<LDL)$.\\
$\alpha_2$&:&  $P(X>UDL)$.\\
$\mu$ &:&  Mean of the distribution.\\
$\mu_{e}$ &:&  Median of the distribution.\\
$\sigma$ &:&  Standard deviation of the distribution.\\
$p$ &:&  Process yield, $i.e$, $\int^{U}_{L}f(x)dx$=$F(U)-F(L)$.\\
$p_{o}$ &:&  Desirable yield, $F(UDL)-F(LDL)=1-\alpha$ with $\alpha=\alpha_1+\alpha_2$.\\
\end{tabular}
\section{Background}
Process Capability analysis is a technique that has application in many segments of the product cycle, including product and process design, vendor sourcing, production or manufacturing planning and manufacturing. Quality of the products has always been a major concern for both consumers and producers. There appears to be a general acceptance of the idea that PCIs can be used only after it has been established that a process is in ``statistical control". This is reasonable, if it simply required that there be no irregular changes in quality level. However, there seems to be, in some quarters, an assumption that the measured characteristic should have a normal distribution, although it is difficult to see why a good industrial process must result in a normal distribution for every measured characteristic. The first PCI was developed by Juran($1974$) and is defined as 
\begin{eqnarray}\label{eq1}
	C_{p}&=&\frac{U-L}{6\sigma}\\&=&\nonumber\frac{d}{3\sigma},
\end{eqnarray}
 where, $d=(U-L)/2$. Note that $C_{p}$ does not depend on process mean. A $C_{p}$ value of $1$ means that $99.73 \%$ of all individual items will be within specification. However, if there is only a slight change in the process mean or a slight increase in the process variation, a larger portion of items will be out of specification. Currently, $C_{p}$ values of $1.33$ or greater are used widely as industry standards. Some companies now require $C_{p}$ values of $1.66$ from their suppliers. A $C_{p}$ of $1.33$ indicates that $99.73 \%$ of products production is within $(100 \%/1.33)=75\%$ of the specification, whereas a $C_{p}$ of $1.66$ shows that almost all product measurements fall within $(100 \%/1.66)=60 \%$ of the specification. For example, if the width of the specification is about $20$ and the standard deviation is $3$,
$$
C_{p}=\frac{30-10}{18}=1.1
$$
indicating that not only can the process produce almost all production within specifications but also $100 \%/1.11=90 \%$ of the specification will contain essentially all the production measurements. Whereas $C_{p}$ values represent the setting in which the process is centered between the specification limits, process are frequently operated with the mean other than halfway between the upper and lower specification limits.\\
When the process mean is off-centered of the specification, the result is that one specification limit (the closer to the process mean) becomes the focal point of the process capability calculation. The modified capability index, represents the situation in which one specification limit is more likely to be exceeded. The $C_{pk}$ is then introduced by Kane($1986$) to reflect the impact of $\mu$ (process mean) on the process capability indices.
\begin{eqnarray}\label{eq2}
	C_{pk}&=& min \left[\frac{U-\mu}{3\sigma}~,\frac{\mu-L}{3\sigma} \right]\\&=&\nonumber\left[\frac{d-|\mu-M|}{3\sigma} \right],
\end{eqnarray}
where, $M=(U+L)/2$. The change in the denominator from six to three standard deviations is the result of the two one-sided quality concerns. For example, the $C_{pk}$ for the situation described above is $C_{pk}= min[0.78, 1.44]=0.78$. When $C_{p}$ and $C_{pk}$ are calculated for a process that is centered, the two calculation methods give the same number. Therefore, $C_{pk}$ is generally preferred because it is not dependent on the process being centered. However, the disadvantage of using only the $C_{pk}$ value is that it does not reveal whether a poor process capability is the result of a process having a large dispersion or an off-centered process.\\
Therefore, both $C_{p}$ and $C_{pk}$ should be used to judge the process capability. If the index values are different, then this is a sign that the process mean is off-center and considerable improvement in the proportion of acceptable product can be made by centering the process mean. $C_{pk}$ has been criticized because it does not measure the process deviation from target. A process may have a high $C_{pk}$ even when the process mean is off-target and close to the specification limits as long as the process spread is small.\\
As an alternative, process capability index, $C_{pm}$, given by Hsiang and Taguchi($1985$), takes into account the influence of the departure of the process mean $\mu$ from the target $T$,
\begin{eqnarray}\label{eq3}
	C_{pm}&=&\frac{U-L}{6\sqrt{{\sigma}^2 +(\mu -T)^2}}\\&=&\nonumber\frac{d}{3\sqrt{{\sigma}^2 +(\mu -T)^2}}\\&=&\nonumber\frac{d}{3\sqrt{E(X -T)^2}}
\end{eqnarray}
where $\mu$ is the process mean and $T$ is the target value and $E(.)$ denotes ``expected value''. It is easy to verify that $C_{pm}$ will possess necessary properties required for assessing process capability. If the process variance( i.e., $\sigma^2$ ) increases(decreases) the $C_{pm}$ will decrease(increase). If the process drifts from its target value (i.e., if $\mu$ moves away from $T$) then $C_{pm}$ decreases. In the case where both the process variance and the process mean change, the $C_{pm}$ index reflects these changes as well. Usually, $T=M$ ; if $T \neq M $ the situation is sometimes described as ``asymmetric tolerances'' (see Boyles ($1994$) and Vannman ($1997b, 1998a$)). The measure $C_{pm}$ sometimes called the ``Taguchi index''. There is also the hybrid index given by Pearn et al. ($1992$)\\
\begin{eqnarray}\label{eq4}
	C_{pmk}&=&min \left[\frac{U-\mu}{3\sqrt{{\sigma}^2 +(\mu -T)^2}}~,\frac{\mu-L}{3\sqrt{{\sigma}^2 +(\mu -T)^2}}\right]\\&=&\nonumber\frac{d-|\mu-M|}{3\sqrt{E(\mu -T)^2}}
\end{eqnarray}
Clearly, $C_{pmk}$ is only meaningful when deviation from target is the main concern. This index is based on the quadratic loss function and, thus, should only be used when there is evidence of a quadratic monetary loss. There are an infinite number of possible loss functions, and in many practical cases, material outside specification limits may result in a total loss rather than a quadratic loss. Clearly $C_{p}\geq C_{pk}\geq C_{pmk}$ and $C_{p}\geq C_{pm}\geq C_{pmk}$. The relation between $C_{pk}$ and $C_{pm}$ is less clear cut. From Equations ($1$) and ($2$) we have \\
Boyles ($1994$) suggested index is
\begin{eqnarray}\label{eq5}
S_{pk}&=&\frac{1}{3}\Phi^{-1}\left[\frac{1}{2}\Phi\left(\frac{U-\mu}{\sigma}\right)+\frac{1}{2}\Phi\left(\frac{\mu-L}{\sigma}\right)\right]~~~~~~~~~~~~~[Boyles, (1994) ]
\end{eqnarray}\\
An enlightening view of relations among our basic PCIs can be obtained from studies of the superstructure PCIs" introduced by Vannman ($1995$). Let $d=(U-L)/2$ be the half length of the specification interval. Vannman($1995$) defined a class of PCIs which is a generalization of the existing indices. It depends on two non-negative parameters $u$ and $v$
\begin{eqnarray}\label{eq6}
	C_{p}(u,v)&=& \left[\frac{d-u}{6\sqrt{{\sigma}^2 +v(\mu -T)^2}}\right]~;~~~~~~~~~~ [Vannman,1995]
\end{eqnarray}
where $\mu$ is the process mean, $\sigma$ is the process standard deviation, $T$ is the target value and $M=(U+L)/2$ is the mid-point of the specification interval. By varying the values of $u$ and $v$, one can easily see that $C_{p}(0,0)=C_{p}$ (Juran ($1974$)), $C_{p}(1,0)=C_{pk}$ (Kane ($1986$)), $C_{p}(0,1)=C_{pm}$ (Hsiang and Taguchi (1985)), and $C_{p}(1,1)=C_{pmk}$ (Pearn, Kotz and Johnson($1992$)). Moreover, when $T=M$ and for $u, v>0$, $C_{p}(u,v)$ has the same interpretation as $C_{pm}$ and $C_{pmk}$.\\
Spiring ($1997$) also defines a PCI, $C_{p}^{(w)}=C_{p}(0,w)$. However, in this definition, $w$ is not necessarily a constant; it may be a function of $\frac{\mu-T}{\sigma}$. In principle this allows $[w(\mu-T)/\sigma]^2$ in Vannman's formula to be replaced by any function of $(\mu-T)/\sigma$. So, in effect,
\begin{eqnarray}\label{eq7}
	C_{p}^{(w)}&=&\frac{C_{p}}{1+g((\mu-T)/\sigma)^2}~;~~~~~~~~~~~~~~~ [Spiring, 1997]
\end{eqnarray}
with a general choice of function $g(.)$, though for practical purposes it should be a positive, increasing function.
\section{PCIs under the assumption of non-normal process distribution}
As already noted, the ``$6$" in Equation $(3.1)$ has been associated with the idea that a normal distribution for $X$ provides a satisfactory approximation. Of course, both practitioners and theoreticians realized that this would not always be the case, and some (at least) of the second group energetically busied themselves with the task of coming up with relevant information and suggestions. Some practitioners, on the other hand, have claimed that $C_{p}$ need not be assessed on the grounds of direct relevance to properties of NC product, though it is not clear what other means of assessment are to be used.\\
At a relatively early date, Clements $(1989)$, in an influential paper, suggested that $``6\sigma"$ be replaced by the length of the interval between the upper and lower $0.135$ percentage points of the distribution of $X$ (this is for a normal N($\mu,~\sigma^2$)). The new PCI is
\begin{eqnarray}\label{eq8}
{C}_{p}^{'}&=&\frac{U-L}{\xi_{1-a}-\xi_{a}}~~~~~~~~~~~~~~~~~~~~~~~~~~~~[Clements, (1989)]
\end{eqnarray}
where $\xi_{a}$ is defined by P[$X\leq \xi_{a}$]=a, taking $a=0.00135$, so that $\xi_{1-a}$, $\xi_{a}$ are the upper and lower $0.135$ percentiles of the distribution of $X$. [For a N($\mu,\sigma^2$) distribution $\xi_{1-a}=\mu+3\sigma,~\xi_{a}=\mu-3\sigma$.]\\
Clements $(1989)$ suggested fitting a Pearson system distribution for $X$, in order to obtain the required $\xi_{a}$ values. Applications of this kind of method, with various assumed distributional forms, have been quite numerous since $1992$. References include: Rodriguez ($1992$), Bittanti et al. ($1998$), Lovera et al. ($1997$) -all Pearson system; Castagliola ($1996$) -Burr distributions; Farnum ($1996$), Polansky et al. ($1998$), Pyzdek ($1992$) -all Johnson system; Padgett and Sengupta ($1996$) -Weibull and log-normal; Mukherjee and Singh ($1997$) -Weibull; Sarkar and Pal ($1997$) -extreme value; Somervile and Montgomery ($1996/7$) -t, gamma, and log-normal; Sundaraiyar ($1996$) -inverse Gaussian. As mentioned above, Polansky ($1998,~2001$) uses a general kernel fitting method.\\
An Index defined on Inter quantile range is given by 
\begin{eqnarray}\label{eq9}
I&=&\frac{U-L}{F^{-1}(1-\alpha_2)-F^{-1}(\alpha_1)}~~~~~~~~~~~~~~~~~~~~~~~~~~~~~~~[Mukherjee, (1995)]
\end{eqnarray}\\
Yeh and Bhattacharya ($1998$) proposed use of a PCI based on the ratios of expected proportion NC to actual observed or estimated proportion NC.
Let $X$ be the random variable associated with the characteristic of a process under study. Here we assume that $X$ is positive and has a continuous distribution. Let $p_{0}$ be the desired proportion of non-conforming output and $p$ be the actual proportion of non-conforming. Then 
\begin{eqnarray}\label{eq10}
a)~~~~~C_{p}&=&\frac{p_{0}}{p}~;~~~~~~~~~~~ [Yeh~and~ Bhattacharya, 1998]
\end{eqnarray}
Yeh and Chen ($1999$) have extended this to multivariate cases. Another PCI, suggested by Yeh and Bhattacharya ($1998$), distinguishes between NC items for which $X$ is less than $L$, and those for which $X$ is greater than $U$.
Let $\alpha_{0}^{L}$ and $\alpha_{0}^{U}$ be the expected proportions of non-conforming products the manufacturer can tolerate on the lower and upper specification limits, respectively. Let $\alpha_{L}=P(X<L)$ and $\alpha_{U}=P(X>U)$ be the actual proportions of non-conformance of the process. The question of whether the process is ``capable" can then be answered by comparing $\alpha_{0}^{L}$ to $\alpha_{L}$ and $\alpha_{0}^{U}$ to $\alpha_{U}$. Thus
\begin{eqnarray}\label{eq11}
b)~~~~~C_{f}&=& min \left[\frac{\alpha_{0}^{L}}{\alpha_{L}}~,\frac{\alpha_{0}^{U}}{\alpha_{U}} \right]~;~~~~~~~~~~~ [Yeh~and~ Bhattacharya, 1998]
\end{eqnarray}
Here the index is directly linked to the probability of non-conformance. The computation involved in estimating this index is more intensive than the conventional PCIs. This perhaps is not a serious constraints in today's computer technology. A new measure of process capability which is in one-to-one correspondence with the process non-conforming fraction $\pi$. Namely 
\begin{eqnarray}\label{eq12}
C&=&\frac{1}{3}\Phi^{-1}\left(1-\frac{\pi}{2}\right)~~~~~~~~~~~~~~~~~~~~~~~~~~~~~~[Borges~and~Ho, (2001)]
\end{eqnarray}
where $\pi=1-P(L\leq X \leq U)$ is the in-control process fraction defective. The index $C$ has some interesting properties. (i) It is one-to-one correspondence with the in-control process fraction defective $\pi$. Processes with the same fraction defective will be equally capable. Moreover, the process capability will respond only to changes in the process fraction defective and not to changes in the distribution of the observed quality characteristic.\\
(ii) If, under process stability, $X$ is normally distributed with mean $\mu=(L+U)/2$ and standard deviation $\sigma$, then 
$$\frac{\pi}{2}=\Phi\left( \frac{L-\mu}{\sigma}\right)=1-\Phi\left(\frac{U-\mu}{\sigma}\right)$$
and
$$C_{p}=\frac{U-\mu}{3\sigma}=\frac{1}{3}\Phi^{-1}(1-\pi/2)=C,$$
i.e., process capability is measured by the $C_{p}$ value of a centered normal process with the same in-control process fraction defective $\pi$. This property makes $C$ values easy to interpret and understand. (iii) It can be applied to discrete and continuous, uni- and multivariate quality indicators and no assumption on the distribution of the observed quality characteristic is made.
\section{PCIs under the assumption of any general process distribution}
Now a PCI is introduced which overcomes the drawbacks of the standard indices discussed earlier. 
It is defined as the ratio
\begin{eqnarray}\label{eq13}
C_{pc}&=&\frac{1-p_{0}}{1-p}~;~~~~~~~~~~~ [Perakis~and~Xekalaki, 2002]
\end{eqnarray}
Where $p_{0}$ denotes the minimum allowable proportion of conformance. $p$ is the process yield lying between $L$ and $U$. The value of $p_{0}$ must be intuitively close to unity and depends on the nature of the examined process and the requirements of the customers. This index is used for process with unilateral as well as bilateral tolerances and can take into account the minimum acceptable process yield(if such quantity has been set).\\
A generalized process capability index given by Maiti et al. (2010), which is
the ratio of proportion of specification conformance (or, process
yield) to proportion of desired (or, natural) conformance.  Almost
all previously defined process capabilities are directly or
indirectly associated with this generalized one.
 \begin{eqnarray}\label{eq3.1}
 C_{py}&=&\frac{F(U)-F(L)}{F(UDL)-F(LDL)}
 \nonumber\\&=&\frac{p}{p_0}.
 \end{eqnarray}
 The following theorem gives different forms of $C_{py}$~when $X$
 has uniform distribution. 
 \begin{t1}\label{th1}
 If $X$ follows uniform distribution over $(a, b), a<b,$ then
 $C_{py}$ reduces to
 \begin{enumerate}
 \item[($a$)] $C_{p}=\frac{U-L}{6\sigma}$~when
 $LDL=\mu-3\sigma$~and $UDL=\mu+3\sigma$ [Juran (1974)]; \item
 [($b$)]$\acute{C}_{p}=\frac{U-L}{\xi_{1-\frac{\alpha}{2}}-\xi_{\frac{\alpha}{2}}}$~when
 $LDL=\xi_{\frac{\alpha}{2}}$~and $UDL=\xi_{1-\frac{\alpha}{2}}$
 with $\kappa=P(X\leq\xi_\kappa)$ [Clements (1989)]; \item [($c$)]
 $I=\frac{U-L}{F^{-1}(1-\alpha_2)-F^{-1}(\alpha_1)}$ when
 $LDL=F^{-1}(\alpha_1)$~and $UDL=F^{-1}(1-\alpha_2)$ [Mukherjee
 (1995)].
 \end{enumerate}
 \end{t1}
 When proportions of desired conformances are $\alpha_1$
 for lower tail and  $1-\alpha_2$ for upper tail, then 
 $C_{py}=\frac{p}{1-\alpha_1-\alpha_2}$, which for a normal process
 with $LDL=\mu-3\sigma$ and $UDL=\mu+3\sigma$ is $C_{py}=\frac{p}{0.9973}$.\\
 \hspace*{0.2in} Theorem $\ref{th1}$ motivates us to formulate the
 generalized index $C_{py}$ given in $(\ref{eq3.1})$.  Let us
 examine the behaviour of the proposed index $C_{py}$ for different
 values of $p$. Obviously, if $p=p_0$ (for normal distribution,
 $p=0.9973$), then $ C_{py}=1$. If the process yield $p$ is greater
 than $p_0$, then $C_{py}>1$  whereas, if $p<p_0$, then $C_{py}<1$,
 and the value of the index approaches zero as p tends to zero.
 Thus, the smallest possible value of $C_{py}$ is zero.\\
 Now the situation is of importance when the
process is off-centered, $i.e$, $F(L)+F(U)\neq1$, but the
proportion of desired conformance is achieved. In that case they
defined the index as follows:
\begin{eqnarray}\label{eq3.2}   C_{pyk}&=&\mbox{min}\left\{\frac{F(U)-F(\mu_e)}{\frac{1}{2}-\alpha_2},~\frac{F(\mu_e)-F(L)}{\frac{1}{2}-\alpha_1}\right\}\nonumber\\&=&\mbox{min}\left\{C_{pyu},~C_{pyl}\right\}
\end{eqnarray}
When $C_{py}$ and $C_{pyk}$ are calculated for a
centered process, they come out to be the same. Therefore,
$C_{pyk}$ is generally preferred because it is not dependent on
the process being centered. If the index values are different,
then this is a sign that the process median (mean in case of
symmetric distribution) is off-centered, and considerable
improvement in the proportion of acceptable
product can be made by centering the process median.\\
\noindent When $\alpha_1=\alpha_2=\frac{\alpha}{2}$, then
\begin{eqnarray*}   C_{pyk}&=&\mbox{min}\left\{\frac{F(U)-\frac{1}{2}}{\frac{1}{2}(1-\alpha)},~\frac{\frac{1}{2}-F(L)}{\frac{1}{2}(1-\alpha)}\right\}\\
&=&\frac{d-|\frac{1}{2}-F(M)|}{\frac{1}{2}(1-\alpha)},
\end{eqnarray*}
where $F(M)=\frac{F(L)+F(U)}{2}$.\\
For a normal process and under the assumption of
uniformity as in Theorem $\ref{th1}$,  $C_{pyk}$ reduces to
\begin{eqnarray*}   C_{pk}&=&\mbox{min}\left\{\frac{U-\mu}{3\sigma},~\frac{\mu-L}{3\sigma}\right\}\\&=&\frac{d-|\mu-M|}{3\sigma}.
\end{eqnarray*}
where $d=\frac{U-L}{2}$.\\
It generally happens that process target $T$ is
such that $F(T)=\frac{F(L)+F(U)}{2}$; if
$F(T)\neq\frac{F(L)+F(U)}{2}$, the situation may be described as
``generalized asymmetric tolerances" [Boyles (1994) and Vannman
(1997,1998) have described by the term ``asymmetric tolerances"
when $T\neq M=\frac{L+U}{2}$]. Under this circumstance, they defined
\begin{eqnarray}\label{eq3.3}
C_{pTk}&=&\mbox{min}\left\{\frac{F(U)-F(T)}{\frac{1}{2}-\alpha_2},~\frac{F(T)-F(L)}{\frac{1}{2}-\alpha_1}\right\}.
\end{eqnarray}
At this point, it would be interesting to note that $C_{pTk}$ is
equal to $C_{pyk}$ when $T=\mu_{e}$.\\
This generalized process capability index overcomes many deficiencies of the PCIs which have been already discussed and more or less all the existing PCIs are directly or indirectly associated with this index. And it can be used for process with unilateral or bilateral tolerances.    
\section{PCIs under the assumption of multivariate process distribution}
A more precise title for this Section would be ``PCIs for Use When $X$ is Multivariate". Many of the PCIs in this group are not, in fact, multivariate. May be they should be, but writers have opted for construction of univariate PCIs, based on the multivariate distributions of $X$. Nevertheless we will term them all ``multivariate PCIs" (MPCIs). References with a title including the word ``multivariate" or ``bivariate" are: Beck and Ester ($1998$); Bernardo and Irony ($1996$); Boyles ($1996b$); Chan et al. ($1991$); Davis et al. ($1992$); Hellmich and Wolf ($1996$); Hubele et al. ($1991$); Karl et al. ($1994$); Li and Lin ($1996$); Mukherjee and Singh ($1994$); Niverthi and Dey ($2000$); Shariari et al. ($1995$); Taam et al. (1993); Tang and Barnett ($1998$); Veevers ($1995,1998,1999$); Wang et al. ($2000$); Wierda ($1992,1993,1994a,1994b$); and Yeh and Chen ($1999$). Multivariate situations are also discussed in the following references, that do not indicate, explicitly, in their titles that this is so: Chan et al. ($1988b$); Wang and Chen ($1998/9$); and Wang and Hubele ($1999,2001$).\\
The univariate specification interval $(L\leq X \leq U)$ is now replaced by a specification region. This may just be constructed from separate specification intervals: one for each variable $X_i$ in $X$. The specification region is then the hyperrectangle
$$\prod_{i=1}^{v}{(L_i\leq X_i \leq U_i)}.$$
However, more complex regions may be used, reflecting perceived relations among the variables in $X$. These are of the general form
 $$L\leq g(X) \leq U.$$ 
Often, $L$ is zero. Possibly for mathematical convenience, $g(X)$ is often taken as a monotonic function of the joint probability density function of $X$. Thus if $X$ is assumed to have a multivariate normal $N_v(\mu,\Sigma)$ distribution, one might take
\begin{eqnarray}\label{eq14}
g(X)=(X-\mu)'\Sigma^{-1}(X-\mu)
\end{eqnarray} 
and regard an item as NC if $g(X)>U$. In this way we obtain the ellipsoidal specification region 
\begin{eqnarray}\label{eq15}
(X-\mu)'\Sigma^{-1} (X-\mu)\leq U.
\end{eqnarray}
An analogue of $C_p$ is 
\begin{eqnarray}\label{eq16}
\frac{\mbox{Volume of} \{(X-\mu)'\Sigma^{-1}(X-\mu)\leq U\}}{\mbox{Volume of} \{(X-\mu)'\Sigma^{-1}(X-\mu)\leq R\}}&=&\left(\frac{U}{R}\right)^v,
\end{eqnarray}
where 
$$P[(X-\mu)'\Sigma^{-1}(X-\mu)\leq R]=1-p.$$
If the distribution of $X$ is multivariate normal then $(X-\mu)'\Sigma^{-1}(X-\mu)$ has a $\chi^2$ distribution with $v$ degrees of freedom and $R=\chi^2_{v,1-p}$ (the upper $100(1-p)\%$ point of the $\chi^2$ or ``chi-squared" distribution with $v$ degrees of freedom).\\
Chen ($1994$) applies this method to the case when the specification region is of the form in Equation $(\ref{eq15})$. The region is defined by
\begin{eqnarray*}
max_{i=1,2,...,v}\left(\frac{X_i-M_i}{d_i}\right)\leq 1
\end{eqnarray*}
with $M_i=(L_i+U_i)/2$ and $d_i=(U_i-L_i)/2$, and Chen defined $MC_P$ as $R^{-1}$, where
\begin{eqnarray}\label{eq17}
P\left[max_{i=1,2,...,v}\left(\frac{X_i-M_i}{d_i}\right)\leq R\right]&=&1-p.
\end{eqnarray}
There can be many variants on these approaches. For example, the $g(X)$ in Equation $(\ref{eq14})$ might be replaced by $(X-\mu)'A^{-1}(X-\mu)$ where $A$ is a positive matrix, not necessarily the variance-covariance matrix of the distribution of $X$.\\
Shariari et al. ($1995$) proposed a truly multivariate MPCI. It contains three components. The first is of the type in Equation $(\ref{eq17})$. The second is the significance level of the Hotelling's $T^2$ statistic
\begin{eqnarray*}
T^2&=&n(\bar{X}-\mu)'S^{-1}(\bar{X}-\mu),
\end{eqnarray*}
which is
\begin{eqnarray*}
P\left[F_{v,n-v}>\frac{n-v}{v(n-1)}T^2\right],
\end{eqnarray*}
where $F_{v,n-v}$ denotes a variable having the $F$ distribution with $v,n-v$ degrees of freedom. The final component just takes values $1$ or $0$ according to a modified process region -defined as the smallest region similar in shape to the specification region, circumscribed about a specified probability contour (of the distribution of $X$) -is or not entirely contained in the specification region.\\
Wang et al. ($2000$) compared this $3$-component MPCI with Chen's ($1994$) $MC_p$ and with an index $MC_{pm}$ proposed by Taam et al. ($1993$) which is also a ratio of two volumes. The volume in the denominator is the same as in Equation $(\ref{eq16})$ with $R=\chi^2_{v,1-p}$ with $p=0.0027$, while in the numerator we have the volume of a ``modified specification region" which is the largest ellipsoid centered at the target that is within the original specification region.\\
Bairamov ($2006$) introduced the conditionally ordered order statistics for multivariate observations. Let $X_{1}$, $X_{2}$, ..., $X_{n}$ $\in S \subseteq R^m$ be i.i.d. random vectors with $m$-variate c.d.f. $F(x)$ and p.d.f. $f(x)$, where $x=~(x_{1},~x_{2},~...,~x_{m})$ and $S$ is the support of $X$. Consider the real-valued function $N(X):R^m \rightarrow R$, $x=~(x_{1},~x_{2},~...,~x_{m})$, which is continuous in its arguments satisfying $N(X)\geq 0$ for all $x\in R^m$ with $N(X)= 0$ if and only if $x=0$, where $0=~(0,~0,~...,~0)$. $N(X_{1})$, $N(X_{2})$, ..., $N(X_{n})$ are i.i.d. random variables with c.d.f. $P(N(X_{i})\leq N(x)),~x\in R$. The function $N(x)$ introduces partial ordering among the random vectors $X_{1}$, $X_{2}$, ..., $X_{n}$ and $X_{1}$ is said to be conditionally less than $X_{2}$ (or $X_{1}$ precedes $X_{2}$) if $N(X_{1})\leq N(X_{2})$. This ordering is denoted by $X_{1}\prec X_{2}$.\\
Selection o
f the function $N(x)$ is independent of the structure of the system and it can be chosen in various ways depending on the conditions available in the process/system. The following selections for $N(x)$ may be of special interest.\\
(a) $N(x_{1},~x_{2},~...,~x_{n})$= $\sum_{i=1}^{n}{a_{i}x_{i}}$.\\
(b) $N(x_{1},~x_{2},~...,~x_{n})$= $min(x_{1},~x_{2},~...,~x_{n})$.\\
(c) $N(x_{1},~x_{2},~...,~x_{n})$= $max(x_{1},~x_{2},~...,~x_{n})$.\\
While the first function reflects the combined effects with the weights $a_{i}$, the last two functions consider the extreme values.\\
For multivariate quality characteristic, if there are lower and upper specification vectors $\underline L$ and $\underline U$ given, then $\underline{L}\prec \underline{X} \prec \underline{U}$ is more meaningful and similarly for lower desirable vector $\underline{LDL}$ (or lower tolerance vector $\underline{LTL}$)and upper desirable vector $\underline{UDL}$ (or upper tolerance vector $\underline{UTL}$). In that circumstance Maiti et al. (2013) defined generalized process capability index for multivariate process quality characteristic as
\begin{eqnarray}\label{}
C_{py}^M&=&\frac{P(\underline{L}\prec \underline{X} \prec \underline{U})}{P(\underline{LDL}\prec \underline{X} \prec \underline{UDL})}\nonumber\\&=&\frac{P(N(\underline{L})< N(\underline{X}) < N(\underline{U}))}{P(N(\underline{LDL}) < N(\underline{X}) < N(\underline{UDL}))}\nonumber\\&=&\frac{F(N(\underline{U}))-F(N(\underline{L}))}{F(N(\underline{UDL}))-F(N(\underline{LDL}))}
\nonumber\\&=&\frac{p}{p_0}
\end{eqnarray} 
In the similar fashion as in the univariate case they defined $C_{pyk}^M$ and $C_{pTk}^M$ as
$$C_{pyk}^M=min\left\{\frac{F(N(\underline{U}))-\frac{1}{2}}{\frac{1}{2}-\alpha_2},~\frac{\frac{1}{2}-F(N(\underline{L}))}{\frac{1}{2}-\alpha_1}\right\}$$ for off-centered process and
$$C_{pTk}^M=min\left\{\frac{F(N(\underline{U}))-F(N(\underline{T}))}{\frac{1}{2}-\alpha_2},~\frac{F(N(\underline{T}))-F(N(\underline{L}))}{\frac{1}{2}-\alpha_1}\right\}$$ for off-target process respectively, where $F(.)$ is the cumulative distribution function of $N(.)$.\\
They described the steps to be followed to calculate process capability index for a set of multivariate quality characteristic data.\\
$1$. Suppose $\underline{X_1}, \underline{X_2}, ..., \underline{X_n}$ are n observation vectors come from a assumed multivariate distribution.\\
$2$. Transform the observations using $N(\underline{X})$. Let $Y_1=N(\underline{X}_1), Y_2=N(\underline{X}_2), ..., Y_n=N(\underline{X}_n)$. Here the distribution function $F(y)=P(N(\underline{X})<y)$ is the structural function associated with $\underline{X}$ via $Y=N(\underline{X})$. Now $N(\underline{X})$ will possesses a distribution.\\
$3$. Now assuming $Y_1, Y_2, ..., Y_n$ are observations, we will find a best fitted distribution for the observations using Q-Q plot technique or any goodness of fit test.\\
$4$. Using this best fitted univariate distribution, PCI is to be calculated following the approach of Maiti et al. (2010).\\
$5$. If there are more than one multivariate distributions assumed and corresponding structural functions are considered, then we will choose that structural function for which the transformed data set is best fitted. Here, to define generalized process capability index for multivariate data, conditional ordering has been used, and multivariate data has been transformed to univariate one using the concept of structural function. After transforming the data, it is as simple as applying generalized process capability indices for univariate data given by Maiti et al. (2010). It can be well understood and comfortably be used by the practitioners.
\section{Conclusions}
 Process capability studies play an important role in the process control since they assist to decide whether a manufacturing process is suitable and the applications meet the necessary quality standards. The assessment of process capability, which now appears to be quite simple, involves significant dimensions in a practical set up. This is due to the fact that some of the conditions necessary to establish process capability not fully satisfiable. These conditions stipulate that the process has to be under control, that the process output has to follow the normal distribution and the observed values of the quality characteristics be statistically independent. Since these conditions are not fully satisfied in many manufacturing situations, the process capability analysis is becoming to be a critical issue. This situation was of course realized by many researchers and resulted in numerous publications in the literature. Therefore, process capability analysis is valid only when the process under investigation is free of any special or assignable causes (i.e., is in-control.).\\
 Recently, the research in the theory and practice of multivariate process capability indices has been very sparse in comparison to the research dealing with the univariate case. At present, for the multivariate capability indices, consistency in the methodology for evaluating this capability is not so much developed. Moreover, it is quite difficult to obtain the relevant statistical properties. Obviously, further investigations in this field are strongly desirable. In further correspondence, we will no doubt provide a fertile growth for new developments in the theory, methodology, statistical properties and applications of the PCIs. 
\section*{References}
\begin{enumerate}
\item Bairamov, I. G. (2006): Progressive type II censored order statistics for multivariate observations, {\it
Journal of Multivariate Analysis}, {\bf 97}, 797-809.

\item Beck, C. and Ester, S. (1998): Multicriteria Capability Indexes, {\it Spektrum}, {\bf 24}, 179-187.

\item Bernardo, J. M. and Irony, T. X. (1996): A General Multivariate Bayesian Process Capability Index, {\it Statistician}, {\bf 45}, 487-502.

\item Bittanti, S., Lovera, M. and Moiraghi, L. (1998): Application of Non-normal Process Capability Indices to Semiconductor Quality Control, {\it IEEE Transactions on Semiconductor Quality Control}, {\bf 11}, 296-302.

\item Boyles, R. A. (1994): Process Capability with Asymmetric Tolerances, {\it Communications in Statistics - Simulation and Computation}, {\bf 23}, 613-643.

\item Borges, W. and Ho, L. L. (2001): A Fraction Defective based Capability Index, {\it Quality and Reliability
Engineering International}, {\bf 17}, 447-458.

\item Boyles, R. A. (1996b): Multivariate Process Analysis With Lattice Data, {\it Technometrics}, {\bf 38}, 37-49.

\item Carr, W. A. (1991): A new process capability index: Parts per million, {\it Quality Progress}, {\bf 24(8)}, 152.

%\item Chan, L. K., Cheng, S. W. and Spiring, F. A. (1991): A Multivariate Measure of Process Capability, {\it International Journal of Modeling and Simulation}, {\bf 11}, 1-6.

\item Chan, L. K., Cheng, S. W. and Spiring, F. A. (1988b): A New Measure of Process Capability, $C_{pm}$, {\it Journal of Quality Technology}, {\bf 20}, 162-175.

\item Chen, H. (1994): A Multivariate Process Capability Index Over a Rectangular Solid Tolerence Zone, {\it Statistica Sinika}, {\bf 4}, 749-758.

\item Castagliola, P. (1996): Evaluation of Non-normal Process Capability Indices using Burr's Distribution, {\it Quality Engineering}, {\bf 8}, 587-593.

\item Choi, B. C. and Owen, D. B. (1990): A Study of a New Process Capability Index, {\it Communications in Statistics - Theory and Methods}, {\bf 19}, 1232-1245.

\item Clements, J. A. (1989): Process Capability Calculations for Non-normal Distributions, {\it Quality Progress}, {\bf 22}, 95-100.

\item Constable, G. K. and Hobbs, J. B. (1992): Small samples and non-normal capability, {\it ASQC Quality Congress Transactions - Nashville}, 37-43.

\item Davis, R. D., Kaminsky, F. C. and Saboo, S. (1992): Process Capability Analysis for Processes With Either a Circular or a Spherical Tolerence Zone, {\it Quality Engineering}, {\bf 5}, 41-54.

\item Farnum, N. R. (1996): Using Johnson Curves to Describe Non-normal Process Data, {\it Quality Enginering}, {\bf 9}, 329-336.

\item Flaig, J. J. (1996): A new approach to process capability analysis, {\it Quality Engineering}, {\bf 9}, 205-211.

\item Juran, J. M. (1974): Juran's Quality Control Handbook, 3rd ed. {\bf McGraw-Hill}, New York, USA.

\item Hellmich, M. and Wolff, H. (1996): A New Approach for Describing and Controlling Process Capability for a Multivariate Process , {\it Proceedings of the ASA Section on Quality and Productivity}, 44-48.

\item Hsiang, T. C. and Taguchi, G. (1985): Tutorial on Quality Control and Assurance - The Taguchi Methods, {\it Joint Meeting of the American Satistical Association}, Las Vegas, Nevada, 188.

\item Hubele, N. F., Shahriari, H. and Cheng, C. S. (1991): A Bivariate Process Capability Vector, {\it Statistical Process Control in Manufacturing edited by J. B. Keats and D. C. Montgomery}, Marcel Dekker, New York, 299-310.

\item Kaminsky, F.C., Dovich, R.A. and Burke, R.J. (1998): Process Capability Indices ; Now and in the Future, {\it Quality Engineering}, {\bf 10}, 445-453.

\item Kane, V. E. (1986): Process Capability Indices, {\it Journal of Quality Technology}, {\bf 18}, 41-52.

\item Karl, D. P., Morisette, J. and Taams, W. (1994): Some Applications of a Multivariate Capability Index in Geometric Dimensioning and Tolerancing, {\it Quality Engineering}, {\bf 6}, 649-665.

\item Kotz, S. and Johnson, N. L. (1993): Process Capability Indices, {\bf Chapman and Hall}, London, U.K.

\item Kotz, S. and Johnson, N. L. (2002): Process Capability Indices - A Review, {\it Journal of Quality Technology}, {\bf 34(1)}, 2-19.

\item Kotz, S. and Lovelace, C. (1998): Introduction to Process Capability Indices, {\bf Arnold}, London, U.K.

\item Latzo, W. J. (1985): Process capability in administrative applications, {\it ASQC Quality Congress Transactions - Baltimore}, 168-173.

\item Li, Y. and Lin, C. (1996): Multivariate $C_p$ Value, {\it Chinese Journal of Applied Probability and Statistics}, {\bf 12}, 132-138.

\item Lovera, C., Moiraghi, L., Montefusco, A. and Valentini, D. (1997): Estimation of Industrial Process Capabilities -Some Proposals and Studies to Overcome the Obstacle of Some Non-normal Distributions, {\it Abstract III, National Congress, SIMAI, Pavia, Italy}, 627-631.

\item Maiti, S. S., Saha, M., Nanda, A. K. (2010): On Generalizing Process Capability Indices, {\it Journal of Quality Technology and Quantitative Management} {\bf 7(3)}:279-300.

\item Mukherjee, S. P. (1995): Process Capability Indices and Associated Inference1	Poblems, {\it Proceedings of the International Conference on tatistical Methods and Statistical Computation}, Seoul, South Korea, 243-249.

\item Mukherjee, S. P. and Singh, N. K. (1994): Sampling Properties of an Estimate of the Bivariate Process Capability Index, {\it Economic Quality Control}, {\bf 9}, 73-78.

\item Mukherjee, S. P. and Singh, N. K. (1997): Sampling Properties of an Estimator of a New Process Capability Index for Weibull Distributed Quality Characteristics, {\it Quality Engineering}, {\bf 10}, 291-294.

\item Niverthi, M. and Dey, D. K. (2000): Multivariate Process Capability - A Bayesian Perspective, {\it Communications in Statistics -Simulation and Computation}, {\bf 29}, 667-687.

\item Padgett, W. J. and Sengupta, A. (1996): Performance of Process Capability Indices for Weibull and Lognormal Distributions or Autoregressive Processes, {\it Internalnational Journal of Reliability, Quality and Safety Engineering}, {\bf 3}, 217-229.

\item Pearn, W. L. and Chen, K. S. (1995): Estimating Process Capability Indices for Non-normal Pearsonian Populations, {\it Quality and Reliability Engineering International}, {\bf 11}, 389-391.

\item Pearn, W. L. and Kotz, S. (1994-95): Application of Clements' Method for Calculating Second- and Third-generation Process Capability Indices for Non-Normal Pearsonian Populations, {\it Quality Engineering}, {\bf 7},
139-145.

\item Pearn, W. L., Kotz, S. and Johnson, N. L. (1992): Distributional and Inferential Properties of Process Capability
Indices, {\it Journal of Quality Technology}, {\bf 24}, 216-231.

\item Perakis, M. and Xekalaki, E. (2002): A Process Capability Index that is based on the Proportion of Conformance, {\it Journal of Statistical Computation and Simulation}, {\bf 72(9)}, 707-718.

\item Perakis, M. and Xekalaki, E. (2005): A Process Capability Index for Discrete Processes, {\it Journal of Statistical Computation and Simulation}, {\bf 75(3)}, 175-187.

\item Polansky, A. M. (1998): A Smooth Non-parametric Approach to Process Capability, {\it Quality and Reliability Engineering International}, {\bf 14}, 43-48.

\item Polansky, A. M., Chou, Y. -M. and Mason, R. L.(1998): Estimating Process Capability Indices for a Truncated Distribution, {\it Quality Engineering}, {\bf 11}, 257-265.

\item Polansky, A. M. (2001): A Smooth Non-parametric Approach to Multivariate Process Capability, {\it Technometrics}, {\bf 43}, 199-211.

\item Pyzdek, T. (1992): Process Capability Analysis Using Personal Computers, {\it Quality Engineering}, {\bf 4}, 419-440.
\item Rodriguez, R. N. (1992): Recent Developments in Process Capability Analysis, {\it Journal of Quality Technology}, {\bf 24}, 176-187.

\item Sarkar, A. and Pal, S. (1997): Estimation of Process Capability Index for Concentricity, {\it Quality Engineering}, {\bf 9}, 665-671.

\item Schneider, H., Pruett, J. and Lagrange, C. (1995): Uses of Process Capability Indices in Supplier Certification  Process, {\it Quality Engineering}, {\bf 8}, 225-235.

\item Shahriari, H., Hubele, N. F. and Lawrence, F. P. (1995): A Multivariate Process Capability Vector, {\it Proceedings of the 4th Industrial Engineering Research Conference}, Nashville, TN, 303-308.

\item Spiring, F. A. (1997): A Unifying Approach to Process Capability Indices, {\it Journal of Quality Technology}, {\bf 29}, 49-58.

\item Spiring, F., Leung, B., Cheng, S. and Yeung, A. (2003): A Bibliography of Process Capability Papers, {\it Quality and Reliability Engineering International}, {\bf 19(5)}, 445-460.

\item Somerville, S. E. and Montgomery, D. C. (1996/7): Process Capability Indices and Non-normal Distributions, {\it Quality Engineering}, {\bf 9}, 305-316.

\item Sundaraiyar, V. H. (1996): Estimation of a Process Capability Index for Inverse Gaussian Distribution, {\it Communications in Statistics -Theory and Methods}, {\bf 25}, 2381-2398.

\item Taam, W., Subbaiah, P and Liddy, J. W. (1993): A Note on Multivariate Capability Indices, {\it Journal of Applied Statistics}, {\bf 20}, 339-351.

\item Tang, P. F. and Barnett, N. S. (1998): Capability Indices for Multivariate Processes, {\it Technical Report}, Division of Computation, Mathematics and Science, Victoria University, Melbourne, Australia.

\item Vannman, K. (1995): A Unified Approach to Capability Indices, {\it Statistica Sinika}, {\bf 5}, 805-820.

\item Vannman, K. (1997): A General class of Capability Indices in the case of Asymmetric Tolerances, {\it Communications in Statistics- Theory and Methods}, {\bf 26}, 2049-2072.

Vannman, K. (1998): Families of Capability Indices for One-sided Specification Limits, {\it Statistics}, {\bf 31}, 43-66.

\item Veevers, A. (1995): A capability Index for Multiple Response, {\it CSIRO Mathematics and Statistics Report DMS 095}, Australia.

\item Veevers, A. (1998): Viability and Capability Indices for Multiresponse Processes, {\it Journal of Applied Statistics}, {\bf 25}, 545-558.

\item  Veevers, A. (1999): ``Capability indices for multiresponse processes'' in statistical process monitoring and optomization edited by S. H. Park and G.G. Vining, Marcel-Decker, New York, NY, 241-256.

\item Wang, F. K., Hubele, N. F., Lawrence, F. P. and Miskulin, J. O. (2000): Comparison of Three Multivariate Process Capability Indices, {\it Journal of Quality Technology}, {\bf 32}, 263-275.

\item Wang, F. K. and Chen, J. C. (1998/9): Caapability Index Using Principal Components Analysis, {\it Quality Engineering}, {\bf 11}, 21-27.

\item Wang, F. K. and Hubele, N. F. (1999): Quality Evaluation Using Geometric Distance Approach, {\it International Journal of Reliability, Quality and Safety Engineering}, {\bf 6}, 139-153.

\item Wang, F. K. and Hubele, N. F. (2001): Quality Evaluation of Geometric Tolerance Regions in Form and Location, {\it Quality Engineering}, {\bf 14}, 203-209.

\item Wierda, S. J. (1992): Multivariate Quality Control Estimation of the Percentage of Good Product, {\it Technical Report, University of Groningen, Groningen}, Netherlands.

\item Wierda, S. J. (1993): A Multivariate Process Capability Index, {\it Transactions of the ASQC Quality Congress}, 342-348.

\item Wierda, S. J. (1994a): Multivariate Statistical Process Control-Recent Results and Directions for Future Research, {\it Statistica Neerlandica}, {\bf 48}, 147-168.

\item Wierda, S. J. (1994b): Multivariate Statistical Process Control, Thesis, Wolters-Noordhof, Groningen, Netherlands.

\item Yeh, F. B. and Bhattacharya, S. (1998): A Robust Process Capability Index, {\it Communications in Statistics - Simulation and Computation}, {\bf 27(2)}, 565-589.

\item Yeh, A. B. and Chen, H. (1999): A Non-parametric Multivariate Process Capability Index, {\it Preprint}. 
\end{enumerate}
\end{document}